\documentclass[graybox]{svmult}

\usepackage{mathptmx}       % selects Times Roman as basic font
\usepackage{helvet}         % selects Helvetica as sans-serif font
\usepackage{courier}        % selects Courier as typewriter font
\usepackage{type1cm}        % activate if the above 3 fonts are
\usepackage{makeidx}         % allows index generation
\usepackage{graphicx}        % standard LaTeX graphics tool
\usepackage{multicol}        % used for the two-column index
\usepackage[bottom]{footmisc}% places footnotes at page bottom
\usepackage{hyperref}
\usepackage{siunitx}

\makeindex             % used for the subject index
\begin{document}

\title*{The NRL Program in X-ray Navigation}
% Use \titlerunning{Short Title} for an abbreviated version of
% your contribution title if the original one is too long
\author{Kent S. Wood, Paul S. Ray}
% Use \authorrunning{Short Title} for an abbreviated version of
% your contribution title if the original one is too long
\institute{Kent S. Wood \at Naval Research Laboratory, Washington, DC USA \email{kentswood@gmail.com}
\and Paul S. Ray \at Naval Research Laboratory, Washington, DC USA
\email{paul.ray@nrl.navy.mil}}

%
% Use the package "url.sty" to avoid
% problems with special characters
% used in your e-mail or web address
%
\maketitle

\abstract{This chapter describes the development of X-ray Navigation at the Naval Research Laboratory (NRL) within its astrophysics research programs. The prospects for applications emerged from early discoveries of X-ray source classes and their properties. Starting around 1988 some NRL X-ray astronomy programs included navigation as one of the motivations. The USA experiment (1999) was the first flight payload with an explicit X-ray navigation theme. Subsequently, NRL has continued to work in this area through participation in DARPA and NASA programs. Throughout, the general concept of X-ray navigation (XRNAV) has been broad enough to encompass many different uses of X-ray source observations for attitude determination, position determination, and timekeeping. Pulsar-based X-ray navigation (XNAV) is a special case.}

%\abstract{Each chapter should be preceded by an abstract (10--15 lines long) that summarizes the content. The abstract will appear \textit{online} at \url{www.SpringerLink.com} and be available with unrestricted access. This allows unregistered users to read the abstract as a teaser for the complete chapter. As a general rule the abstracts will not appear in the printed version of your book unless it is the style of your particular book or that of the series to which your book belongs.\newline\indent
%Please use the 'starred' version of the new Springer \texttt{abstract} command for typesetting the text of the online abstracts (cf. source file of this chapter template \texttt{abstract}) and include them with the source files of your manuscript. Use the plain \texttt{abstract} command if the abstract is also to appear in the printed version of the book.}

\section{Introduction}

The Development of X-ray navigation at the Naval Research Laboratory (NRL), occurred within the broader context of the Laboratory's program in UV, X-ray, and $\gamma$-ray astrophysics. Astrophysics preceded applications, and the navigational applications began to appear explicitly in internal programmatics only after about 1988. Navigation, construed broadly, is the subject of this chapter but it cannot be covered without reference to the enabling science. X-ray navigation derives entirely from knowledge accumulated since World War II, in contrast to centuries of heritage for navigation based on optical methods.

X-rays do not penetrate the Earth's atmosphere below the ionosphere, hence the onset of the development of X-ray astronomy and its applications was delayed until there were methods of reaching high altitudes. Balloons notwithstanding, it was the advent of the rocket and satellite that instituted changes. Through more than a half century, NRL participated in instrument system design, test and flight, and pursued basic and applied research regarding the X-ray sky. Results that could potentially be exploited included the distinctive variability signatures of the sources as well as the short wavelengths and penetrating nature of the X-rays themselves. The pace of progress is measured by the facts that 13 years elapsed from discovery of solar X-rays (1949), to the first extra-solar source (1962) and then another three decades elapsed to the first detection of an X-ray millisecond pulsar (MSP), in 1993. By that time, the understanding of source classes and their variability signatures had matured. Accumulating knowledge of the X-ray sky was paralleled by major changes in how application opportunities were perceived. When X-ray MSPs were found it did affect a major change in programmatic direction, but investment continued on other approaches. MSP aspects are well covered in other chapters --- this one covers complementary non-MSP aspects, both for intrinsic interest and because they offer solutions to current challenges. Strict historical order will often be sacrificed to logical flow of the development of X-ray navigation, but in actuality that flow largely parallels chronology.

Section 2 treats the beginnings, introducing principal lines of investigation. Section 3 covers times when X-ray sky populations were catalogued and considers how celestial X-ray populations pointed to applications. Section 4 covers thinking that led to NRL's Unconventional Stellar Aspect Experiment, and beyond. Section 5 continues those themes into an era of NASA facilities, particularly \textit{RXTE} and \textit{NICER/SEXTANT}. The abbreviation ``XNAV'' has come to be identified with X-ray navigation accomplished with rotation-powered pulsars, primarily MSPs. This abbreviation will be reserved for pulsar-based navigation. To reinforce a recurring distinction, a different abbreviation, ``XRNAV'' will be used for the broader concept of which XNAV is a special case.

\section{Beginnings through 1960s}
\label{sec:1.0}

NRL's X-ray navigation has been characterized by these major themes:

\begin{enumerate}
\item Exploiting persistent variability of sources, including pulsations of all types, but also other rapid variability signatures, and
\item Transitions, meaning eclipses and occultations, which serve to establish sharp shadow edges for angular resolution and define long, fiducial straight lines in space, with diffraction effects minimized.
\end{enumerate}

%\\*\indent {(i) Exploiting persistent variability of sources, including pulsations of all types, but also other rapid variability signatures, and}
%\\*\indent {(ii)Transitions, meaning eclipses and occultations, which serve to establish sharp shadow edges for angular resolution and define long, fiducial straight lines in space, with diffraction effects minimized.}

Starting in the 1940s, NRL was a major player in the earliest stages of the US space program, following an earlier phase of high-altitude research on the ionosphere. This was prior to establishment of NASA. Initially, space research was done with sensor payloads on sounding rockets; however, beginning with the \textit{SOLRAD} program, satellite platforms for sensors replaced rockets. Interest in astronomy evolved out of solar pursuits. NRL researchers led by H. Friedman made first detection of the Sun in X-rays in 1949 \cite{1}. This revealed a solar X-ray luminosity high enough to explain the day-side ionosphere. Optical spectroscopy of coronal lines had led to theoretical inference of plasma hot enough to emit X-rays. Modeling the ionosphere had led E. O. Hulburt at NRL to estimate the expected level of solar X-ray flux. However there was little understanding of how this hot plasma was maintained and how the corona could be hotter than the photosphere. The first direct observation established the magnitude of the Solar X-ray flux. Variability during flares became the focus of research. This work approximately established the X-ray luminosity expected for coronae of other stars, but it was below sensitivity thresholds of instruments then available. 

More luminous neutron star and black hole X-ray source classes were largely unexpected, even in theory, prior to serendipitous discovery of Scorpius X-1, in a rocket flight by Giacconi et al. in 1962 \cite{2}, although Friedman had once speculated the Crab Nebula might be detectable. Once Sco X-1 was discovered another twenty years were required to establish firmly its neutron star nature, but the discovery flight was followed immediately by further sounding rocket flights from NRL and elsewhere. A picture of galactic plus some extragalactic sources emerged. A review written in 1966 \cite{3} provides a historical snapshot of those early impressions of the X-ray sky, gained using rockets. The nomenclature for many of the brighter X-ray sources by constellation (Sco X-1, Her X-1, Cyg X-1, etc.) is heritage of this era. 

From solar flare work, Friedman and collaborators were accustomed to encountering X-ray variability. Comparing early rocket flights they found that the source Cyg X-1, which today is recognized to be an accreting black hole, had varied significantly in luminosity. This was among the earliest clues that dramatic source variability would be a striking difference between the X-ray sky and the familiar optical sky. NRL found the second brightest source, the Crab Nebula (called Taurus X-1 for a while) \cite{4} and began to specialize in understanding that source. A measurement central to the second theme above (transitions) was NRL's search for a neutron star in the Crab Nebula using a lunar occultation on 7 July 1964 \cite{5}, several years before it was known that neutron stars could be pulsars. The occultation probed the source angular structure without using X-ray optics. In 1967 radio pulsars were discovered, which motivated a different search using fast timing of individual X-ray events An NRL rocket in 1968 found the first X-ray pulsar, the Crab Pulsar \cite{6}. Pulsed emission at X-ray wavelengths was surprising. 

The two principal themes in the Chapter derive from these two flights. The 1964 occultation flight marked first use of methods based on edges and transitions. The 1968 flight provided the first X-ray pulsar detection, discovering the brightest pulsar known in the X-ray sky, still a crucial resource for X-ray navigation. 

\subsection{Crab Occultation Flight}
\label{sec:1.1}

In 1964, near the dawn of X-ray astronomy, a lunar occultation was the best way to achieve fine angular resolution. The scientific goal was to determine whether the X-ray source, then known for only a year, was a point source (neutron star) or extended (the Nebula). The extent of the Crab Nebula was known in other wavelengths but unknown in X-rays. The inertial space direction of the Moon as seen from Earth advances 360 degrees in a month, which equates to about a half arc-second per second. The Crab Nebula extends about an arc minute in X-rays (a fraction of its extent in visible wavelengths) so that the central portion can be covered (or uncovered) by the Moon in a time within reach of a rocket flight, which affords only a few minutes of observation. An X-ray source extending of order an arc-minute would disappear gradually as the Moon passed between Earth and the Nebula, but a point source would show an abrupt drop at ingress. 

\begin{figure}
\includegraphics[width=4.5in]{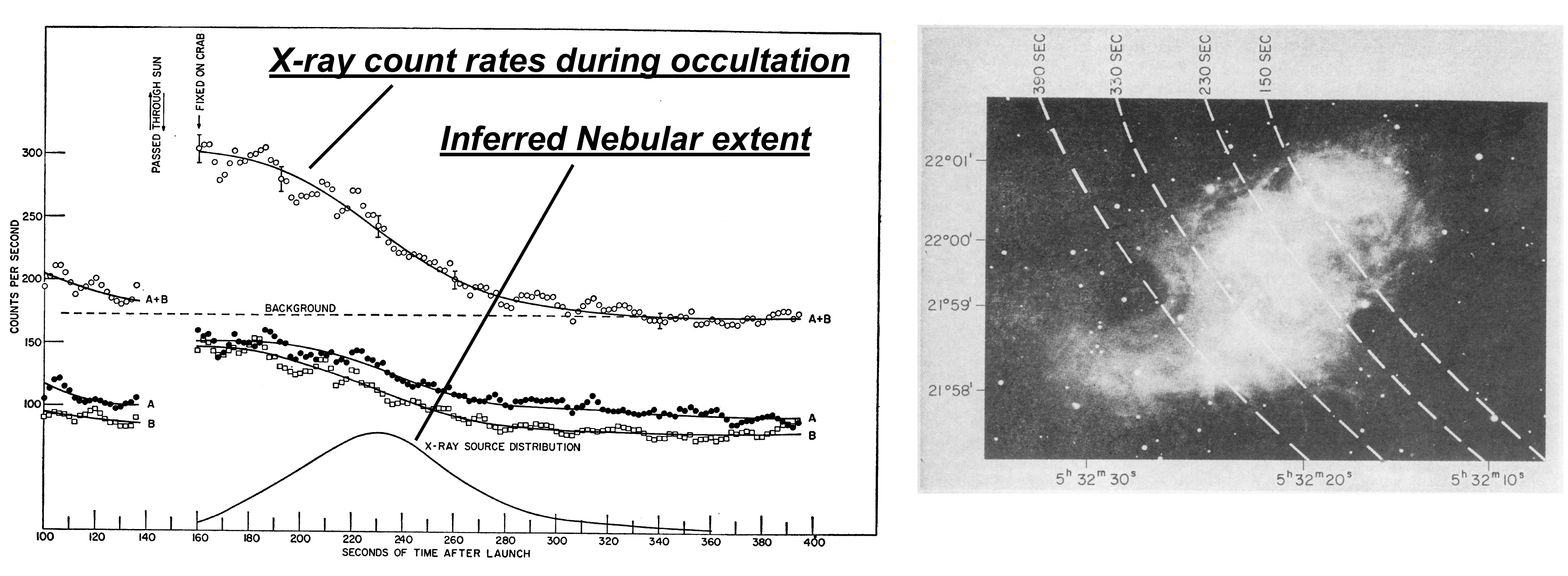}
\caption{(a) For the 7 July 1964 flight \cite{5}, left panel shows the count rates from the two geiger counters, the combined rate, and the Nebula projected profile derived from the derivative; (b) right panel shows times during flight represented as advancing stages of the lunar limb, allowing X-ray brightness distribution inferred from the occultation to be cross-referenced to optical brightness distribution of the Crab Nebula.  Compare this panel with Figure 13, which depicts a more recent observation.}
\end{figure}

In 1964, there was speculation covering possible outcomes for the occultation observation, ranging from a point source to an extended Nebula, or combinations. Neutron star models were based on thermal emission from a hot surface, since pulsars were not yet known. Estimates existed for the neutron star size and surface temperature but there was great uncertainty regarding these numbers. 

The 7 July 1964 flight was a classic experiment in the effective use of a sounding rocket, one where perfect launch timing was essential, but it was severely limited by detector areas. The detectors were two Geiger counters, each with 114 cm$^2$ of frontal area. The flight time history is shown in Figure 1a. The occultation date was fixed by the lunar orbit. In early July the Sun is near the Crab Nebula, hence it happened that the instrument system scanned through the Sun on its way to acquiring the Nebula. Then it was able to watch the Nebula fade as the lunar limb advanced. Launch time had to consider the inertial space coordinates of the Nebula centroid and location of lunar shadow, not on the ground but on the rocket trajectory at altitude. The combination of lunar limb and rocket trajectory gave an apparent limb advance rate of $\sim$1 arc-second per second as shown on Figure 1b. The (projected) nebular brightness distribution was obtained from the derivative of the count rate history, registered on the sky relative to the Nebula using the lunar ephemeris and rocket trajectory. 

In this way analysis established the angular extent of the Crab Nebula in X-rays but only placed limits on X-rays from a neutron star at the position of the optical object then referred to as Baade's Star. This was an example of navigation being used for X-ray astronomy, rather than the other way around. Knowledge of the rocket trajectory, both predictively in planning the moment of the launch and \textit{post facto} in data analysis, was essential. The fact that the derived nebular brightness centroid roughly coincides with the optical centroid vindicates the decision to time the launch so as to capture the interval when the Moon was covering the central Nebula.

\subsection{Crab Pulsar Discovery Flight}
\label{sec:1.2}

The Crab Pulsar lies near the Nebula centroid and by 1968 it was known to pulse in optical wavelengths. Its spin-down is the energy supply for the total Nebula. It exemplifies a rotation-powered pulsar (RPP) and a pulsar wind Nebula (PWN), as distinguished from pulsars powered by accretion or magnetic field annihilation. X-ray pulsations provide only about 5 \% of the total flux from the source in keV energies, accounting for why the occultation method had not detected it, phase-averaged, in the 1964 flight. The X-ray discovery \cite{6} established that, in this pulsar, the main and secondary pulses were coincident across radio, optical, and X-ray wavelengths, with little offset in pulse phase. Light curve features coincided from 10$^8$ Hz to at least 10$^{17}$ Hz. Later work extends the upper bound another 10 orders of magnitude. 

Concurrently with these developments, Roger Easton, often called the ``Father of GPS,'' was undertaking Project Vanguard, the \textit{Timation} satellites and other efforts leading to GPS concept, which occurred in stages during the years 1955 -- 64 \cite{7}. The use of periodic signals for navigation was thus a longstanding NRL interest, dating to this time. To this point the chronology concerns times before the first author was at NRL and derives from the published literature and conversations with H. Friedman, T. Chubb, E. Byram, J. Meekins, G. Fritz,  and P. Wilhelm, all of whom were at NRL during the 1960s. 

\section{Late 1960s ---1970s: Classes, Variability Signatures, and Uses}
\label{sec:2.0}

The discovery of Scorpius X-1 \cite{2} and other sources in subsequent rocket flights in the 1960s gave an initial picture of the X-ray sidereal sky down to some tens of milliCrabs (a convenient if vague unit, signifying ~1/1000th of the flux of the Crab Nebula). The diversity of these first detections raised questions as to what source populations existed and their properties. Papers, both observational and theoretical, discussed the make-up of the X-ray sky, always pointing to the need for better surveys. Subsequent surveys performed using the \textit{\textit{Uhuru}} (1972), \textit{Ariel V} (1975), and \textit{HEAO-1}(1977) satellites pushed the limiting flux down to ~1 milliCrab, with the total count of known sources approaching 1000. They revealed further population diversity and refined the characterization of prominent source
classes, obtaining their spectral and temporal properties, and their angular extent, if any.

It was being learned the X-ray sky has no steady point sources --- all sources were either extended or variable. Understanding the various classes of galactic and extragalactic sources encountered and developing models for the physics underlying their X-ray emission occupied a generation of astronomers. Variability often provided crucial information for choosing preferred models and establishing intrinsic parameters of sources. A portfolio of variability signatures emerged from this international effort. In this era, prior to X-ray optics, the principal paths to clarification of variability were flying larger apertures or staying pointed with 3-axis stabilization on single sources to accumulate observing time, or both

\begin{figure}
\includegraphics[width=4.5in]{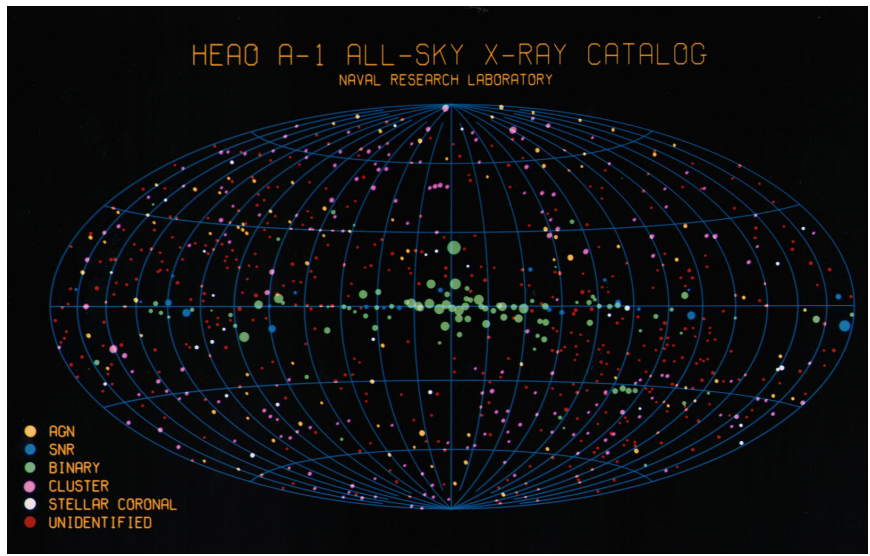}
\caption{All-sky map, in galactic coordinates, of the \textit{HEAO} A-1 X-ray source catalog, based on the first impartial sky survey in energies 0.25-25 keV \cite{8}. The legend at left gives top-level breakdown by class. The "SNR" (supernova remnant) and "Cluster" (cluster of galaxies) classes are extended, hence less useful for navigation.  The remaining classes are variable point sources.  Dot size for a source is logarithmic in flux. Sco X-1 is just above the Galactic Center, while the Crab Nebula is at the far right (color-coded as SNR, despite the pulsar within the Nebula).}
\end{figure}

Figure 2 shows the all-sky catalog \cite{8} produced by NRL from the \textit{HEAO} A-1 Experiment, which flew from 1977 -- 79. Although this was not the first sky survey, it was the first in which the survey was conducted impartially, sweeping out successive sky segments daily along meridians of Ecliptic longitude until the entire sky was covered. Over the mission life three sweeps of the sky were completed. This remains qualitatively a fair representation of the brightest ~1000 sources, with the important reservation that some depicted sources have faded and others appeared with passage of time. There is a continuing turnover in the bright sources that presents both a hazard and an opportunity for X-ray navigation.	

By the early 1970s it was firmly established that neutron stars were X-ray sources. Periodicities were known, not only for the Crab Pulsar but others such as Her X-1 and Cen X-3. The Fourth \textit{Uhuru} Catalog \cite{9} listed nine accreting X-ray pulsars (SMC X-1, X Per, Vela X-1, Cen X-3, 4U 1223-62, GX 304-1, 4U1626-67, Her X-1, and GX1+4), for some of which periods had been found using instruments other than \textit{Uhuru} itself. Save for SMC X-1, the spin periods were longer than 1 second. The list kept growing with subsequent surveys. Rotation-powered and accretion-powered pulsars were recognized as distinct classes, with different mechanisms for producing X-ray pulses. For all source classes, timing signatures were being characterized and modeled. The millisecond domain was opened by the Crab pulsar detection in 1969; its main pulse turned on and off in a little over a millisecond. In the era of sounding rockets, the black hole source Cygnus X-1 had been found variable, showing aperiodic variability, over decades in timescale, down to milliseconds. 

Global characterization of the X-ray sky is the essential preliminary to any practical application of source properties. MSPs were not yet established as an X-ray source class. The first MSP, PSR B1937+21, was discovered in radio only in 1982 \cite{10}. The X-ray program at NRL was partly justified as studying backgrounds relevant to potential future space systems. To the extent that direct utilization of source properties was considered it was customary to ask why one would perform a given task in X-rays rather than, say, optical wavelengths. Potential responses included (i) advantageous detector system options, (ii) exploitation of distinctive variability signatures of X-ray sources, and (iii) short wavelengths, \textit{i.e.}, minimal diffraction. Sometimes pursuit of these questions revealed difficulties. For example accreting X-ray pulsars such as Cen X-3 were found to have considerable variation in their spin periods, because of torques applied by accreting material supplied from binary companions. They were poor clocks, but practical utilization does not necessarily require precise periodicity (see relative navigation, below).

On one occasion, \textit{HEAO-1} used a lunar occultation to characterize emission from a cluster of galaxies. \textit{Uhuru} found eclipses in the accreting pulsar, Hercules X-1. \textit{HEAO} A-1 found the first unpulsed but eclipsing low-mass X-ray binary (LMXB), the system now known as X 1658-298 \cite{11,12}. Eclipses provide a different kind of periodic signal, infrequent but with a distinctive, sharp edge. The precision of this celestial clock was initially unknown, but the timescale for evolution of the orbit was known to be long, of order 10$^8$ yr. This and another eclipsing system, Exo 0748-676, were monitored for years, and will be discussed further in later sections. Through the 1980s, the Crab remained the best short-period X-ray clock and sharp edges in binary eclipses the best longer-period clocks, where cycle count could not be lost by drifts.

\section{1980s -- 2006: XLA, USA, and X-ray Navigation}
\label{sec:3.0}
The crucial years from ~1980 to 2006 will now be described in detail. We start with an overview of these decades, during which the navigational potential of X-ray astronomy came to be pursued seriously at NRL. The Laboratory's role encourages evaluating engineering spinoffs of basic research. Navigation using X-rays was pursued along several lines, from time variability (including periodicity) of sources, to their extreme angular compactness, to technical advantages specific to X-rays, and coupled with emerging technologies available for detection of X-rays from space. The obvious GPS analogy meant attention was paid to periodic signals, but transitions and sharp X-ray shadows provided another avenue. There was a practical connection between navigation and onboard computation. At this time radiation-induced faults were a hindrance to computing in space. Consequently while the NRL group was building a navigation payload it included a testbed for reliable space-based computing and autonomy. Programmatically, efforts in these areas advanced in parallel from the end of \textit{HEAO-1} (1979) through the USA experiment (launched 1999 on the \textit{ARGOS} satellite). Two satellite facilities, USA and \textit{Ginga}, plus analysis done for a third facility that remained conceptual (the X-ray Large Array, XLA) provided a programmatic framework through the year 2000. Then, building upon USA, a project specific to pulsar-based navigation, called ``XNAV,'' was undertaken with DARPA. For several years XNAV was coupled with attempting to realize a follow-on to USA, which halted once it became likely that \textit{NICER/SEXTANT} would fly. The end of the DARPA program (2006) closed this phase.

\subsection{XLA to USA}
\label{sec:3.1}

After \textit{HEAO-1}(March, 1979), NRL specialized further in X-ray source variability. An idea from the early 1980s was a large (100 m$^2$) array called the X-ray Large Array (XLA), which had a NASA Pre-Phase A study, but never flew. It was proposed to be attached to the Space Station, then a future endeavor in planning. (For the XLA retrospective see \cite{13}. ) Similar large arrays continue to be proposed today to NASA and ESA but the largest X-ray apertures actually flown have yet to exceed 0.5 m$^2$. The science enabled uniquely by a large array (apertures from several up to several tens of square meters) without optics is not directly germane to navigation and will not be described. The immediate fallout of efforts on XLA was that assured benefits of large exposure (defined as area x time product) on X-ray sources could be had, for certain purposes, by a descoped approach to a smaller aperture, using repeated observation on prioritized sources to achieve large exposure. That descope path from XLA evolved into the much smaller (area 0.2 m$^2$) Unconventional Stellar Aspect (USA) Experiment. What pertains here from XLA consists of analysis studies with bearing on particular aspects of navigation, analyses concerning capabilities achievable only with large arrays, not small systems.

\subsection{XLA: Occultations for Ultra-fine Angular Resolution}
\label{sec:3.2}
XLA motivated analysis of using occultations ambitiously to achieve ultra-fine angular resolution. A catalog of X-ray sources, coupled with astrophysics of identified populations, implies the angular scales at which sources of various types should become resolvable. For some classes (e.g., large X-ray coronal structures inferred in RS CVn-type systems), sources would become resolvable if X-ray angular resolution reached $\sim$100 milli-arcseconds. However, for accreting supermassive black holes micro-arcseconds were clearly needed. In the 1980s X-ray focusing optics systems and X-ray coded aperture systems were both striving toward finer angular resolution. It is hard for focusing systems to reach angular resolution $<$100 milli-arcseconds (m.a.s.), save perhaps in combination with X-ray interferometry \cite{14}. However a large array combined with occultation can reach m.a.s. scales without focusing, combining geometric optics with the short X-ray wavelength. The angular performance improves linearly as the array area grows, \textit{i.e}., the resolution limit scales as (area)$^{-1}$, not as (area)$^{-1/2}$. This is true for fairly bright sources able to stand out against background, but the diversity of populations among bright X-ray sources assures reasonable representatives of most classes.

Reaching ever-finer scales is ultimately limited by X-ray diffraction from the occultor edge.  XLA's large aperture could reach the m.a.s. domain, starting with the Moon as occultor, which would give access to Ecliptic latitudes of $\pm$6 degrees over 9 years, roughly 10 \% of full sky. This portion includes the Crab Nebula, 3C273 and other important sources. Thus far, this is astrophysics, not astronautics, but it leads directly back to astronautics if one wishes to access the other 90 \% of the sky. An artificial occultor in high orbit would increase sky access to 4$\pi$  sr(100 \% ). 

Wood and Breakwell \cite{15} analyzed an artificial occultor for use with large arrays in low Earth orbit( LEO), based on a steerable, smooth edge, able to access the entire sky for high angular resolution on targets. Specifically it envisioned a delta-wing occultor placed in a high orbit whose radius would equal the mean radius of the lunar orbit, but with its orbit plane perpendicular to the lunar orbit plane.  (See Figure 3; dates appearing on the figure were wishful thinking in 1987.)  A basic engineering requirement is to have the edge advance slowly enough to bring out angular structure at a statistically significant level. X-ray diffraction does not limit performance, but another requirement is to be able to know the occultor position accurately and steer it onto the target.

\begin{figure}
\includegraphics[width=4.5in]{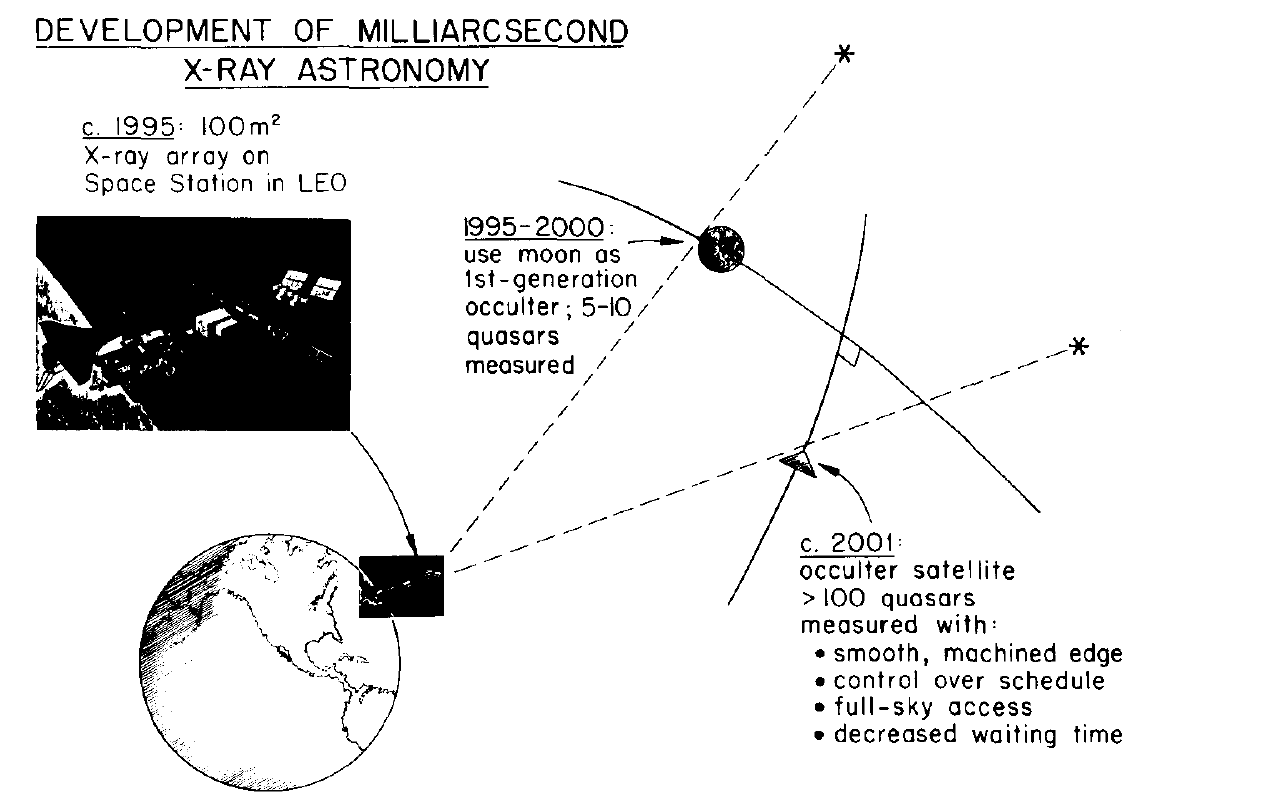}
\caption{Proposed orbital geometry for artificial occultor, from \cite{15}.  Using thrusters and positional information from GPS satellites on the far side of the earth, the upper occultor platform is navigated to produce artificial occultations to be observed by the detector array located on the International Space Station.  Detection of the occultation at the intended time validates the precision navigation exercise. }
\end{figure}

Accurate orbit knowledge and control requirements in such a high orbit were addressed in the thesis of C.M. Roitmayr \cite{16}. An occultor in LEO was rejected for limited performance and excessive control authority requirements. At the distance of the Moon the occultor could be steered to alignments to produce artificial occultations, using the gravitational perturbations of the Sun and Moon plus thrusters. For navigating the high occultor it was necessary to know where it was. The answer was to use GPS navigation based on transmitters from the farther side of the Earth. GPS transmitters are beamed toward nadir, but with an opening cone angle that exceeds the angle subtended by the Earth. 

The navigation concept was reduced to a Kalman filter formulation, and fueling and command strategies were worked. This was another exercise in using navigation to achieve an X-ray astronomy objective. However it could also be viewed as an example of constructing long straight lines in space for fiducial or other purposes, to scales of meters over more than 100,000 km. While the artificial occultor has never flown, the idea remains viable. Progress made with \textit{RXTE} is described in Section 5.5.

\begin{figure}
\includegraphics[width=4.5in]{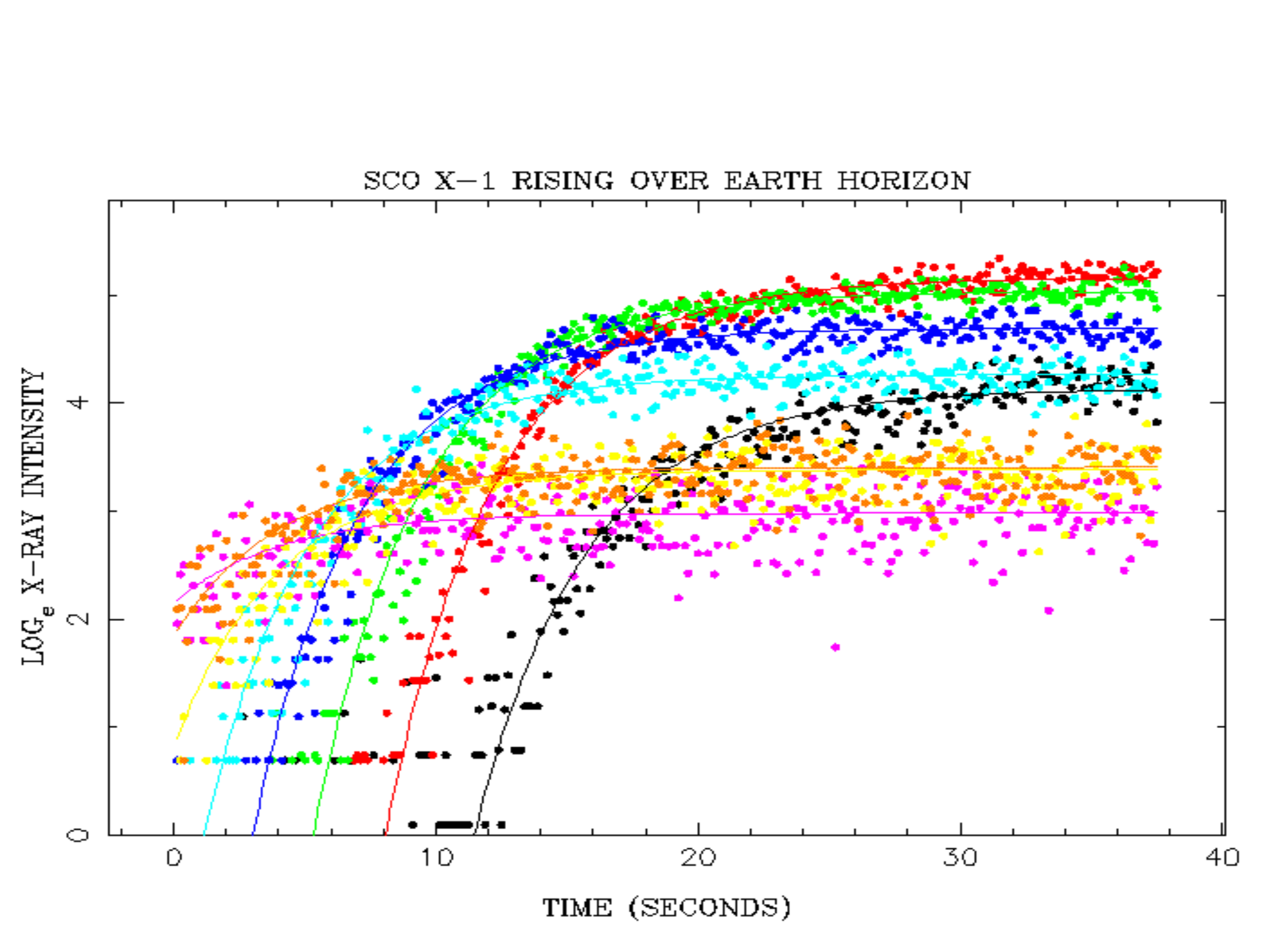}
\caption{Sco X-1 rising through the Earth's atmosphere as observed with a proportional counter array on the Japanese satellite \textit {Ginga}. Higher energies emerge earlier, \textit{i.e.}, energy, color-coded runs (lowest ot highest) black, red, green, blue, cyan, yellow, magenta. The energy dependence of this type of transition is displayed differently (for a different occultation event) in Figure 8.  A series of such occultations can yield a precise determination of orbit elements near a planet or moon. }
\end{figure}

Consider now Earth occultations, a special case of eclipses, \textit{i.e.}, by the atmosphere when source transitions at Earth limb, as applied to the problem of determining a satellite position when it is in low orbit near a planet. Atmospheric occultation profiles are rounded, not sharp-edged, but do yield acceptable navigational information, conveniently referenced to the Earth. It is necessary to model the atmosphere (as well as planetary shape, e.g., Earth oblateness) to fit occultation curves and derive orbits. This technique is not confined to X-rays and has been used in other wavelengths as well. It becomes a trade study of costs and performance benefits. 

NRL research on X-ray limb sensing for navigation began with the Japanese satellite \textit{Ginga} (1987-1991), used by NRL scientists as guest investigators. \textit{Ginga} was typically scheduled to eliminate Earth occultations, \textit{i.e.}, it acquired sources after they were estimated to be above the Earth limb and ceased observing before they dipped below Earth limb. Nevertheless occultations could occur unintentionally. A rising transit of Sco X-1 in 1989 (Figure 4) provided inspiration to the development of USA. This transit would not have been seen had \textit{Ginga} been in the orbit used for scheduling. The orbit had changed and the exact time of transit measured the change of orbital elements, \textit{i..e.}, provided useful navigational information. This methodology was incorporated into the USA mission concept.

\subsection{Earth Occultations as X-ray Atmospheric Diagnostics}
\label{sec:3.3}
Earth occultations can work another way, providing atmospheric diagnostics for remote sensing. In this application one (ideally) knows the source celestial coordinates accurately and also knows the state vector of the observing satellite, whereupon the atmospheric density profile can be treated as the unknown. Rounded occultation profiles seen when X-ray sources set into the atmosphere were regarded, correctly, as degrading the precision of navigational information, but not beyond usefulness. Good understanding of the structure and composition of the atmosphere would largely compensate for this, providing suitable templates for fitting transition profiles. The X-ray observing system might be used for atmospheric research to develop the templates. Earth occultation work that began with Ginga continued with USA and came to full fruition in the \textit{RXTE} era (section 5.3).

\begin{figure}
\includegraphics[width=4.5in]{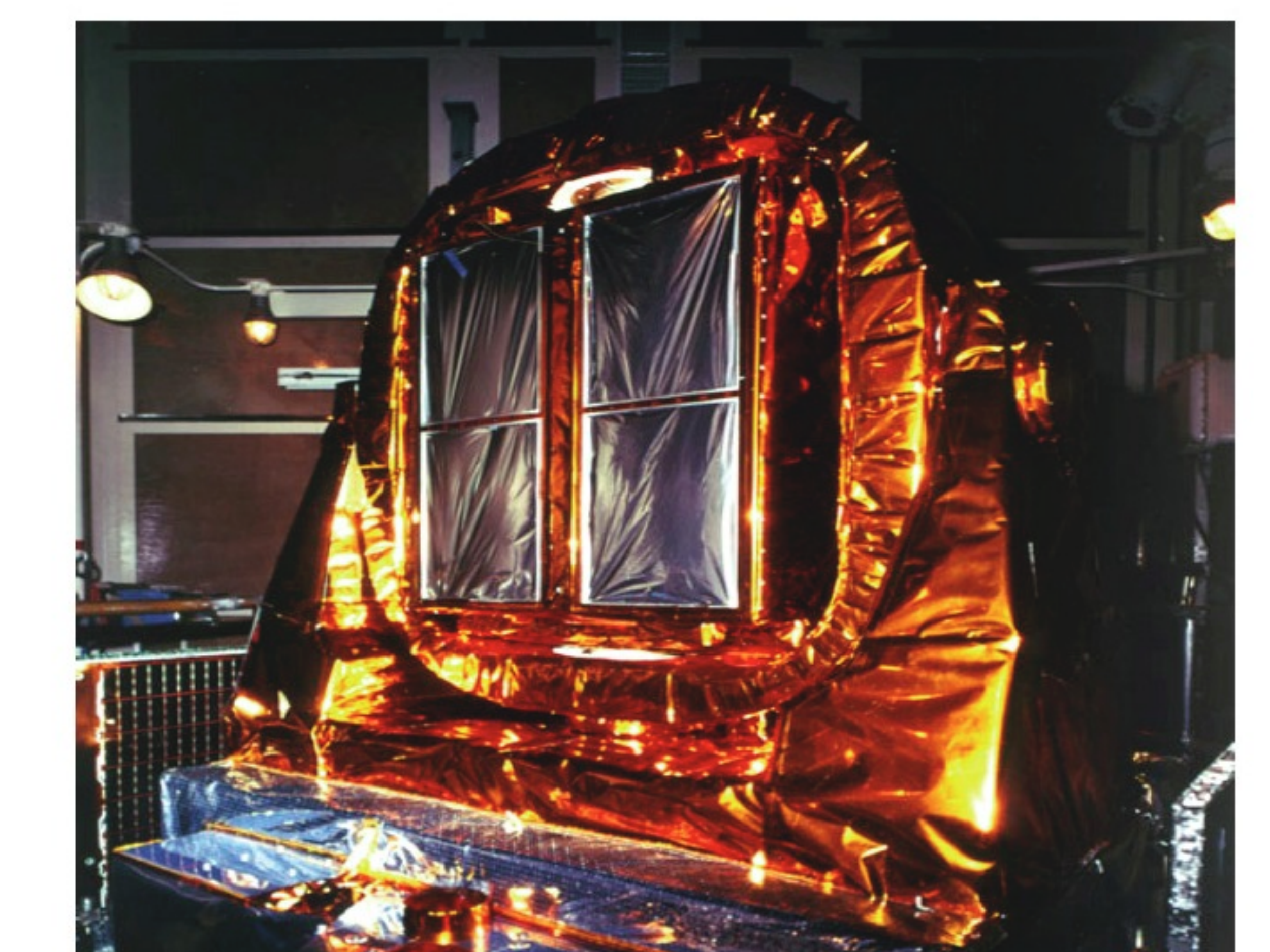}
\caption{A photograph of the USA experiment, the first flight experiment in X-ray navigation, during testing at NRL. The silver-colored panels are heat shields over the proportional counters. The surrounding frame (gold) is the pylon and 2-axis gimbal for offset pointing from \textit{ARGOS}.  The central electronics box, including the computing testbed described in the text, resides near the base of the pylon.}
\end{figure}

By 1988 it was clear that any large-aperture experiment such as XLA would be slow in realization. Attention turned to achieving useful results calling for high exposure with small payloads. This led to the Unconventional Stellar Aspect (USA) experiment, developed from conception in 1988 to launch in 1999. USA was a multi-purpose instrument. It was the first flight in the field of X-ray navigation, and also a technology study for autonomous operation. Autonomy takes several meanings, but in this context it meant ability use a sensor to gather situational information, process it onboard to obtain a state vector, and apply that result as needed.

The USA experiment consisted of two large proportional counters mounted in a 2-axis gimbal (see the instrument paper, \cite{17}, for full details), and attached to the \textit{ARGOS} satellite flown by the U. S. Air Force.

Two papers published during the USA design phase surveyed X-ray navigation \cite{18, 19}. Aspects covered were attitude determination, position determination and timekeeping, with methods used including pulsars, occultations, and eclipses. In the early 1990s, interest in X-ray navigation was for near-Earth use, not interplanetary navigation. The most favorable options for each navigational task were described, judging by achievable accuracy. For timekeeping the best clocks included the Crab and an eclipsing binary. The Crab pulsar TOAs could be predicted to milliseconds over months. For position determination in earth orbit, the best option consisted of timing Earth occultations of sources, either rising or setting. A variant on the \textit{Ginga} transition depicted in Figure 4 (summing over energies and not dividing counts into energy bands) was used to, show how in-track position could be determined from a single transition to an accuracy $\sim$350 meters; an orbit solution could be refined to tens of meters by combining occultations over a few orbital periods. This respectable accuracy, comparable to what is expected for contemporary systems using MSPs, is possible because the transition on a \textit{bright} source can be timed, by fitting the sharp rise, to tens of milliseconds. That precision multiplied by the orbital velocity ($\sim$7 km/s) gives the positional accuracy. In comparison, a determination of phase (TOA) is converted to positional accuracy multiplying by the velocity of light. The slower orbital velocity gives advantage to the occultation method by a factor of c/v$_{orbital}$ or $\sim$4 x10$^4$ for low Earth orbit, though that advantage may be partly cancelled if the pulsar phase can be determined to much better timing accuracy than the Earth occultation, possible only with an X-ray MSP. In 1993, without MSPs, occultations easily gave better accuracy than any known pulsar, but required that the instrument track X-ray sources to the Earth horizon. USA's gimbal was accordingly so designed. The occultation method is at its best near a planet, which is also just where the pulsar method is at disadvantage. An obvious path for future development is to meld the two approaches into more general solutions. (See Secgtion 6.1.)

USA's first look at X-ray navigation was thus in the broad (XRNAV) sense. USA as a whole was an exploratory avionics package based on X-ray sensors. The USA proposal was submitted to the U.S. Air Force Space Test Program in 1988. During 1988 -- 92 its progress was tracked mainly in internal DoD documents while it remained in an approval process, revised and reviewed yearly by the Space Experiment Review Board. Even the 1988 version addressed XRNAV. USA was formally described in publications only when matched with a launch opportunity, which also began to freeze the flight configuration, and led to a funding line. NRL was the primary center of activity. Stanford University and the Stanford Linear Accelerator Center then entered into partnership for production.

USA supported development of X-ray navigation other ways as well. Among publications describing USA during the pre-launch phase was the first Ph.D thesis on X-ray navigation \cite{20}, by J. Hanson. It addressed use of X-rays for attitude determination and included discussion of pulsars, describing a phase-locked loop utilizing the recorded pulsation.  

\subsection{Autonomy and Space-Based Computation}
\label{sec:3.4}
Reliable (processor fault-free) onboard computation was recognized as essential to navigation. The USA experiment therefore incorporated a testbed (Advanced Space Computing and Autonomy Testbed  --- ASCAT) for computing, believed the first of its kind. Being able to compute without errors in the space environment is also essential to any autonomy. It compared performance of two strategies for reliable computing in in the fault-inducing radiation environment of space, namely hardware radiation-hardening of processors vs. fault-tolerance software. Two processors were flown, one utilizing the hardware strategies and the other implementing the software approaches, then they were compared as to kinds and frequencies of faults. With the software methods the less expensive commercial processor was able to achieve tolerable performance. Also, the hardened processor was not completely immune to faults. Complete presentations of ASCAT testbed results are found in \cite{21, 22}. The computer testbed was able to receive the X-ray event data stream onboard.

\subsection{USA as First X-ray Star Tracker}
\label{sec:3.5}

\begin{figure}
\includegraphics[width=4.5in]{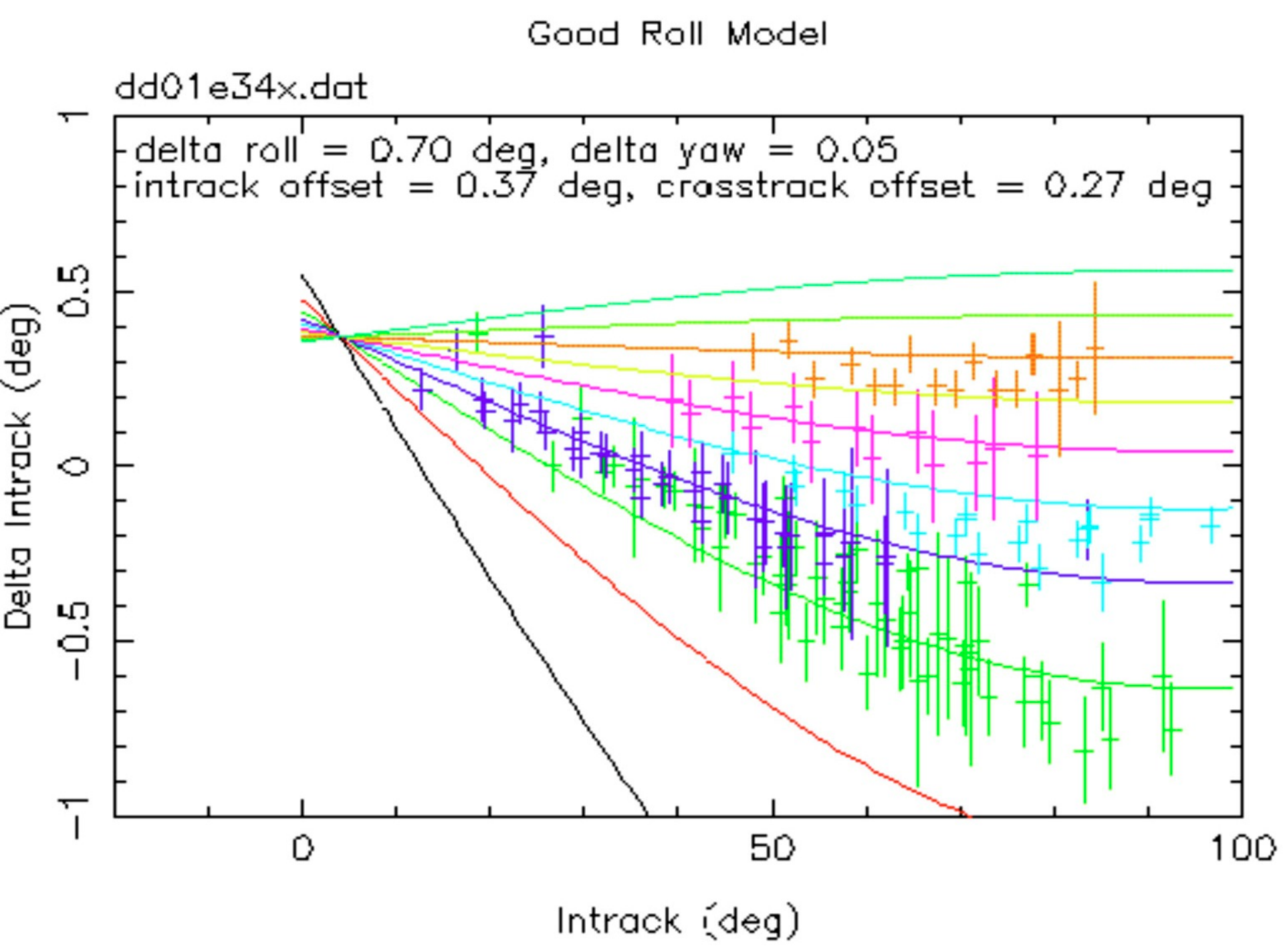}
\caption{Data analysis from scans of a point source, used to analyze \textit{ARGOS} pointing \cite{23}. Colored lines fit the points of the same color solely because a pointing offset of the instrument model is correctly modeled as an error in roll.  A best fit to an assumed yaw error  (not shown) would yield unacceptable fits.  This was used during flight to diagnose an aspect problem with the satellite, hence it constitutes feedback from an X-ray instrument to aspect control. }
\end{figure}

In one flight experiment, USA was used as an X-ray star-tracker to diagnose and correct a problem with the \textit{ARGOS} spacecraft attitude control. The satellite had been missing targets and it was unclear whether it was missing them in roll or yaw (pitch being unlikely). The problem was successfully diagnosed using the USA instrument in a scanning mode similar to what had been envisioned in Hanson's thesis, though complicated by the equatorial mounting and the nadir-pointed attitude of the satellite. This may be the first use of an X-ray sensor as input to attitude control of a flying satellite. (See \cite{23}.) Despite availability of the space-based computing testbed, the feedback loop for this exercise went via data analysis on the ground, thus it was not an autonomy demonstration.

\subsection{USA and Pulsars: Position and timekeeping}
\label{sec:3.6}

Millisecond X-ray pulsars (MSPs) were detected using \textit{Rosat} \cite{24} while USA was in construction. Implications for X-ray navigation were recognized immediately, as was the fact that the sources were at the limit of what USA might hope to detect. The bright MSP PSR B1821-24 was added to the USA observing plan (to explore time transfer as well as navigation), however MSPs proved too faint to be easily isolated by USA. Detection required phasing up the pulsar over a protracted period in the \textit{ARGOS} orbit. For faint sources, an X-ray optical system is needed to support source isolation from background. For USA, the Crab Pulsar was the principal target for pulsar timing applications. Difficulties were encountered in calibrating its phase accurately but it was eventually accomplished. Results on the Crab Pulsar were presented at astronomical conferences and published in engineering discussions \cite{25, 26, 27}. The approximate level of timekeeping demonstrated was a fraction of the width of the main pulse of the Crab. Because of its brightness the Crab Pulsar remains important for XNAV. It will be used along with MSPs in SEXTANT. USA also detected one of the so-called anomalous X-ray pulsars (AXPs), 4U 0142+61. This is a magnetar with fairly stable spin period, $\sim$8.7 s. The navigational significance was that it was much fainter than the Crab; its long period facilitated detection. Whether this class of neutron stars fills a need in X-ray navigation remains to be seen. They might provide coarse information, for example for cold start. Unlike eclipsing binaries, these pulsars do not turn off.

\subsection{Pulsar Navigation Patent Deriving from USA Program}
\label{sec:3.7}

USA led to a patent for pulsar-based X-ray navigation \cite{28}, covering pulsed celestial sources in general as well as the cold-start problem and other technical aspects. The broader XRNAV concept was by now difficult to patent because of our own earlier papers from the 1990s. A University of Maryland engineering student, S. Sheikh, collaborated on USA analysis and contributed to the patent while completing his thesis \cite{29} in 2005, the first thesis on MSP-based XNAV. Concurrently, there were further papers on this \cite{30,31}. The latter discussed replacing the proportional counters used in USA with silicon pixel devices, a concept already under active consideration for a further STP flight.  (Additional contributions by Sheikh are covered in another chapter.)

\subsection{The Silicon X-ray Imager (SIXI)}
\label{sec:3.8}

The Silicon X-ray Imager (SIXI) was proposed to the Space Test Program as a follow-on to USA, while USA was still under construction. Like USA, it was submitted for annual evaluation by the Space Experiment Review Board, starting in 1993. Like USA, it made explicit reference to XRNAV, advancing beyond the capabilities of USA. The use of ``Silicon'' in the name signified abandonment of proportional counter technology in favor of solid-state detectors, which were then being explored for flight use in the aftermath of improvements and cost reductions realized in the early 1990s. Borrowing from XLA, SIXI was proposed to be mounted to the \textit{International Space Station}, (\textit{ISS} ) which by then had a stable design and implementation plan and was approaching first launches. Although this was not true in 1993, a specific experiment interface called the Express Pallet was defined during the years SIXI remained in review and SIXI proposed to use it. NICER now uses this same interface. SIXI was equipped with a coded aperture that would enable it to perform better than USA as an X-ray star tracker. SIXI was designed to do offset pointing at celestial targets from the Express Pallet, the same concept now being used in \textit{NICER/SEXTANT}. SIXI never flew. The \textit{NICER/SEXTANT} approach emerged as a preferable alternative. 

A mature version of SIXI was described in \cite{32}. This article provided a update on X-ray navigation, including discussion of using MSPs.

\subsection{The XNAV Program}
\label{sec:3.9}
Concurrently, DARPA ran a program called XNAV to explore practical approaches to pulsar-based X-ray navigation. These ideas were associated with the possibility of a demonstration with a payload attached to the \textit{ISS}. Some of the details of the proposed SIXI went into the initial announcement of opportunity from DARPA. Two design approaches were eventually selected and evaluated competitively using a pulsar simulator. NRL, LANL, Ball Aerospace, and NASA GSFC participated. Designs with and without focusing optics were worked out in detail, subjected to systems analysis, and compared as to performance. The advantages of optics for isolating faint millisecond pulsars became decisive. Further improvements along these lines later led to \textit{NICER/SEXTANT}. The XNAV proposal was October 2004 and the program ran through 2005 and 2006. The end of the DARPA program effectively marks the end of this phase of the development of X-ray navigation. 

\subsection{Concurrent Efforts: \textit{Fermi} LAT and MSPs}
\label{sec:3.10}
During these same years the NRL group was simultaneously working on the \textit{Fermi} Large Area Telescope (LAT), designed for high-eneergy $\gamma$-rays. The silicon sensor approaches proposed for SIXI and studied under the DARPA program derived in part from experience in the design of the tracker for the \textit{Fermi} LAT. The space-based computing experience from USA was applied to development of the onboard data acquisition system of the LAT. Following launch (2008) the \textit{Fermi} LAT would make many discoveries of new gamma-ray pulsars, including MSPs. The accompanying chapter by P. Ray \cite{33} covers this development. 

\section{2001 -- 2011: Studies during \textit{RXTE} era}
\label{sec:4.0}
Following the end of USA in 2000, no further X-ray astronomy mission with a navigation theme was flown or manifested before \textit{NICER/SEXTANT}, however the Rossi X-ray Timing Explorer (\textit{RXTE}) (launched 1995; end 2012) permitted progress to be made on aspects of XRNAV. \textit{RXTE}, like USA, lacked the X-ray optics needed to see very faint sources in presence of background, but had a larger aperture and its low orbital inclination made backgrounds benign in comparison with USA's. It was thus able to pursue topics beyond USA, including successful X-ray millisecond pulsar observations. With the limited time devoted to these sources all that was possible was source characterization, meaning characterization of invariant characteristics such as light curve and spectrum. The bright MSP, PSR B1821-24 was observed for $\sim$200 ks. Over a suitable integration it is possible to characterize the measured phase of the MSP by a TOA referenced to a radio measurement, and correcting for the satellite orbit. The TOA may in principle be referenced to either Earth center or the Solar System barycenter, provided the same is done for the radio ephemeris. After calibrating and correcting for any intrinsic radio-to-X-ray phase offset, the difference between a radio ephemeris and a representative X-ray TOAs (in seconds) can be multiplied by the velocity of light to obtain a measure of how accurately the X-ray TOA could be used for navigation. A preliminary analysis of the \textit{RXTE} TOA for PSR B1821-24 \cite{34} showed the level then achievable was between 30 and 300 km, \textit{ i.e.}, the TOA was accurate to about 0.1 ms.  See also \cite{35}.

\textit{RXTE} was not able to go beyond this and conduct any demonstration of pulsar navigation. Crucial enabling resources and capabilities that will be available on SEXTANT were missing. These include (i) a software system (on board) that accumulates the pulsar light curve using (ii) a propagated orbit also available onboard, and (iii) signal to noise sufficient to see pulsations and measure the TOA before the error envelope in the onboard propagated orbit grows too large to support phasing up the X-ray counts. \textit{RXTE} data could be used in a \textit{post facto} orbit determination on the ground to see the pulsations, and this information could be used in certain kinds of feasibility calculations.

\subsection{\textit{RXTE} Used for Non-MSP X-ray Applications}
\label{sec:4.1}
More pertinent contributions of \textit{RXTE} lay in broader XRNAV topical areas, specifically: (i) unsurpassed eclipse studies, principally on the source Exo 0748-676, (ii) the most detailed study to date of the use of Earth occultations for atmospheric diagnostics, (iii) study of statistical limitations of ``relative X-ray navigation,'' meaning use of rapid X-ray variability not characterized by periodicity, and (iv) the best lunar occultation results yet achieved on the Crab Nebula and Pulsar, in which the intentions of the 1964 flight were fully realized with high signal-to-noise.

\subsection{Eclipse Timing Noise in Exo 0748-676}
\label{sec:4.2}
By now we know of about a dozen eclipsing LMXBs but in the RXTE era only two were clearly established, those being X1658-298 and Exo 0748-676. The former turned on for a brief period around 2000, but it was not enough for any major advances to be made in characterizing it. Exo 0748-676 was observed for many years through many eclipses. The nearly square signature of the eclipse, dropping at ingress and rising again at egress, could be fitted with a template to derive a best estimate of the mid-eclipse time. This technique was developed and applied in a series of papers. Instead of showing a quiet system with high-accuracy clock-like periodicity, quite noticeable departures were found. They were of a noisy character, not representable with a low-order polynomial fit and not predictable.

\begin{figure}
\includegraphics[width=4.5in]{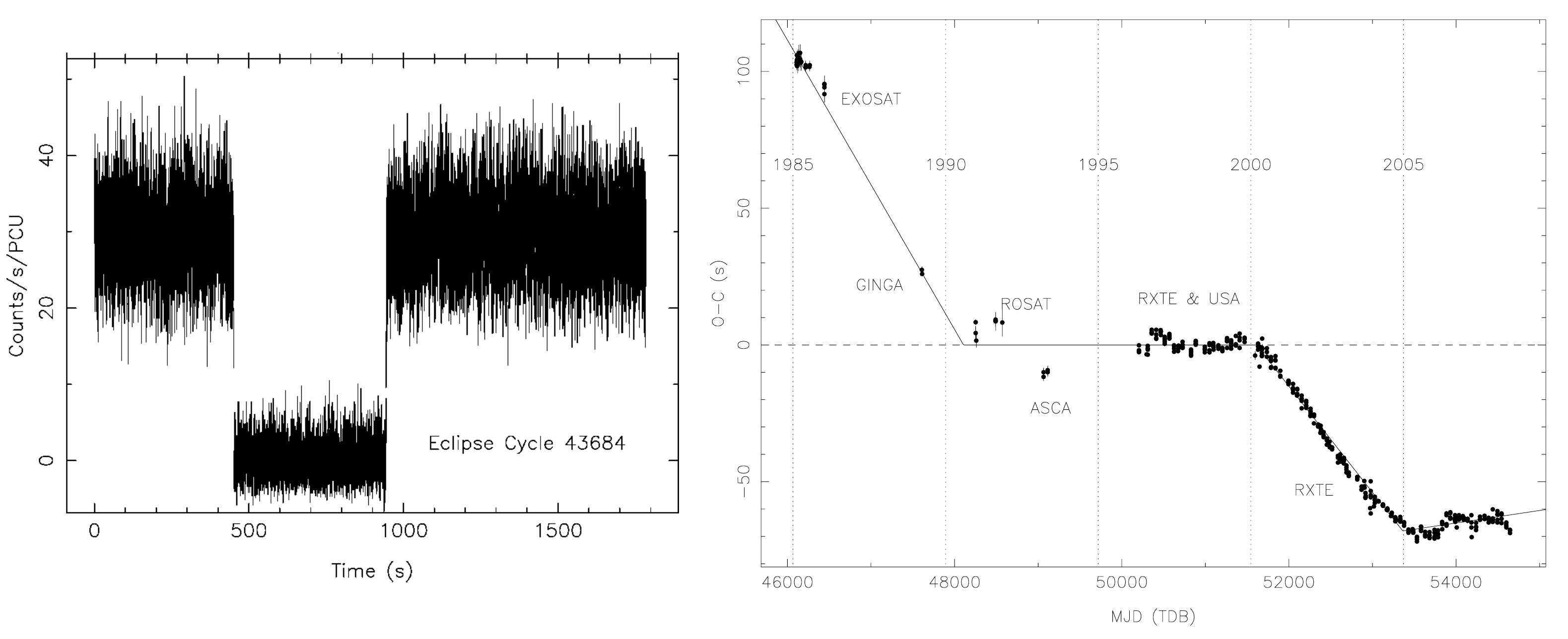}
\caption{(a) A single eclipse from Exo 0748-676 observed with \textit{RXTE}. The sharp edges at ingress and egress are fitted to yield a precise time of mid-eclipse. (b) Advance of mid-eclipse time with epoch, showing timing irregularities discussed in the text. }
\end{figure}

This discovery had somewhat negative implications for using LMXB eclipse events as markers for absolute timekeeping. A mitigating consideration being researched today is that there seems to be a bound on the cumulative buildup of phase drift. Cycle count is not lost over many years, but modeling the short term excursions, for example as polynomial fits, has no demonstrated predictive value. To predict eclipses across decades the mean period is sufficient, and furthermore better than high-order polynomial fits to excursions, which can diverge if extrapolated. Understanding the excursions in terms of underlying mechanism remains a worthwhile challenge and may eventually bring benefits to applied use. Several simple models for the variations have been excluded. One that seems to offer promise is that angular momentum is alternately deposited in and removed from the quadrupole moment of the companion star. This model had been developed for other types of binary sources and was adapted to this instance \cite{36, 37}. By now it is being applied to another class of X-ray MSPs, the redback class, which transitions between MSP and LMXB behavior \cite{38, 39}.

For this analysis, usable eclipses were not limited to \textit{RXTE} alone but instead went earlier to \textit{Exosat}, included some from \textit{Rosat} and USA, and then a higher frequency of eclipse observations once the source had been approved as a candidate for frequent revisit by \textit{RXTE}. The series did not go through the full life of \textit{RXTE} because while \textit{RXTE} was still flying the source went into quiescence in October 2008. Currently (2015) the picture of these orbital period variations is receiving new data from a further outburst.

\subsection{X-ray Diagnostics Applied to the Earth's Atmosphere}
\label{sec:4.3}

Originally, \textit{RXTE} followed the practice of acquiring sources after they were above Earth horizon and leaving them before they set. Because of this policy, it observed no Earth horizon crossings. However after USA had established the value of observing these transitions a proposal was made to continue with RXTE to obtain better profiles. The prime source used was the Crab Nebula. The study combined Earth occultations from USA with those of \textit{RXTE}, on the Crab and Cygnus X-2 \cite{40}. This demonstrated the ability to extract atmospheric density. When X-ray Earth occultations are used to probe the atmosphere they are most sensitive in the altitudes from 80 to 150 km, for typical observing energies and the Earth atmosphere density profile, which lies in the thermosphere. Models predict effects associated with global climate change deriving from buildup of greenhouse gases at the thermopause, at ~ 90 km altitude \cite{41}.

\begin{figure}
\includegraphics[width=4.5in]{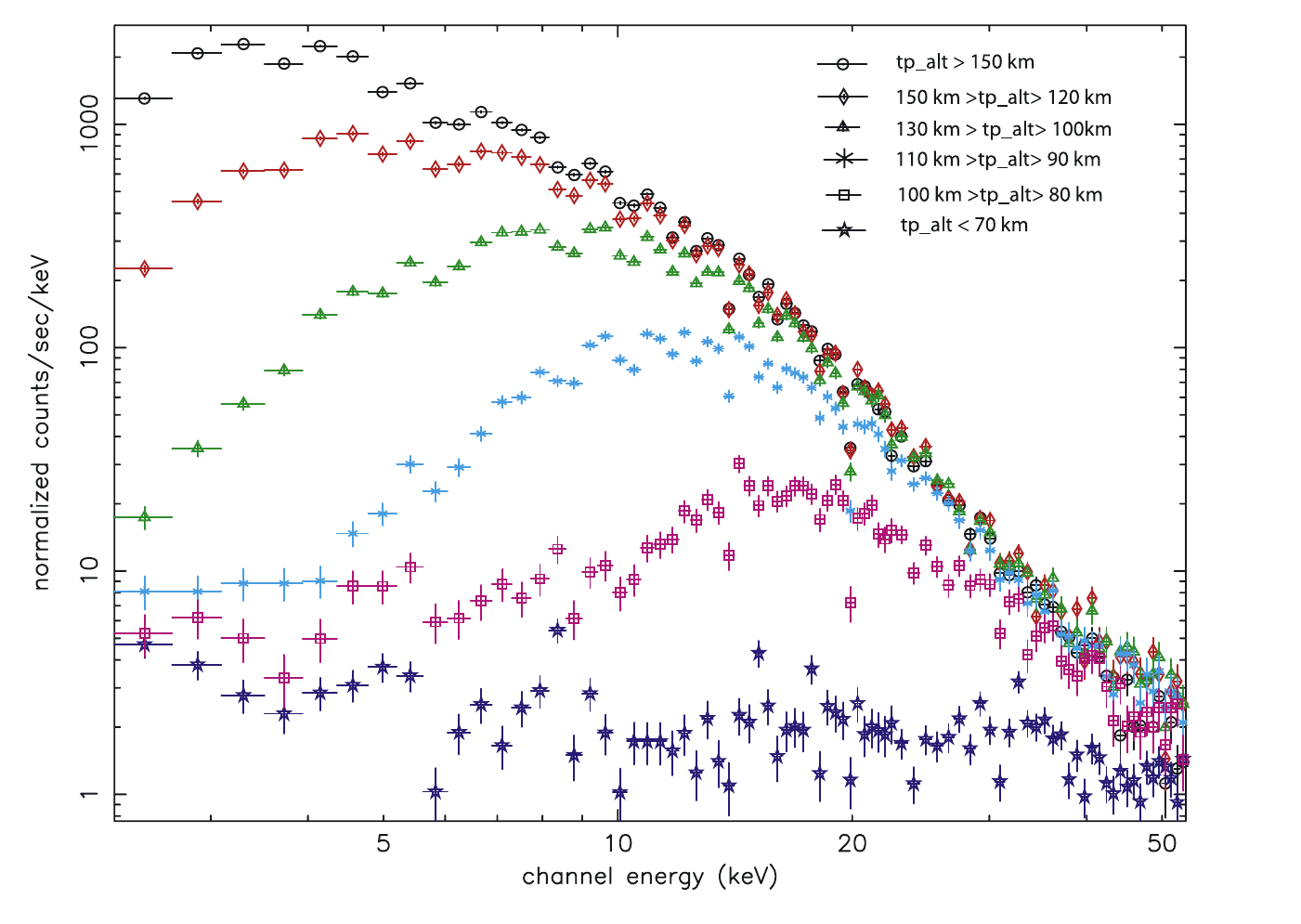}
\caption{An Earth occultation of the Crab, showing energy spectra in \textit{RXTE} proportional counters at successive times as the source sets, from \cite{40}.  Spectra are labeled by tangent altitude.  The highest is fully un-attenuated and the lowest  fully attenuated, \textit{i.e.}, no atmospheric transmission remaining in the energy band.  Compare this depiction of an Earth occultation with Figure 4. }
\end{figure}

While greenhouse gas buildup raises temperature at sea level it has the opposite effect above the thermopause by enhancing radiative cooling. The cooling should cause settling, modifying density profiles at higher altitudes. This must be distinguished from solar effects, by monitoring over times longer than the solar cycle. The \textit{RXTE} occultation study, covering epochs to ~2002, provides a baseline for future comparisons. The atmospheric response (from this effect alone and not combining with solar cycle modulation) has already been modeled. The problem for the future is reduced to establishing requirements on observation systems and monitoring schedules. Besides environmental monitoring this would maintain a current atmospheric model to use for occultation-based navigation in the evolving Earth atmosphere.

\subsection{Relative Navigation by Cross Correlation}
\label{sec:4.4}

\begin{figure}
\includegraphics[width= 4.5in]{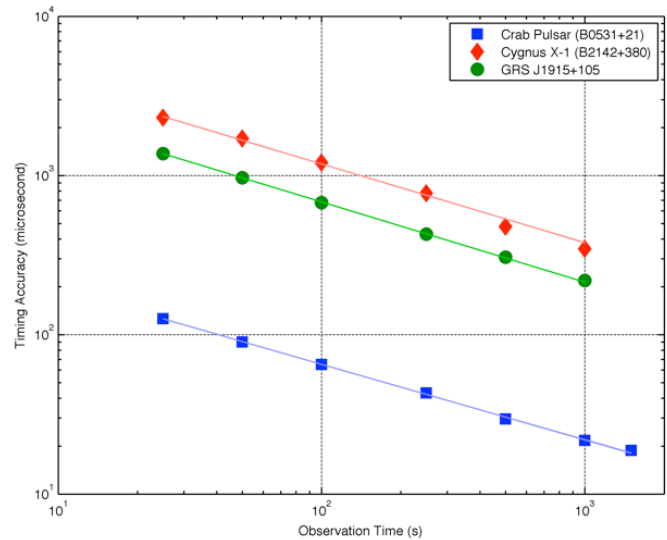}
\caption{X-ray Relative Navigation.  This is a simulation made using flight data of achievable performance using three different celestial sources.  The sources are all bright  (among the brightest sources shown in Figure 2) and the display shows achievable accuracy as a function of integration time. }
\end{figure}

In X-ray relative navigation, two X-ray detectors on two satellites observe the same source and use cross-correlation of their data streams to measure their relative displacement. If one has a well-determined state vector then the relative navigation can be used to assign a state vector to the other. Another mode of utilization is that if both satellites have acceptable position knowledge then the relative offset can be used for time transfer. If the objective is merely to determine the achievable accuracy, which is what was done with \textit{RXTE}, then this distinction is not very important. 

\textit{RXTE} provided only a single satellite but had multiple detectors in its array and did thus provide independent data streams that could be cross correlated to ascertain how accurately relative navigation could be done in principle. Sources used for this purpose were the Crab Pulsar and two black hole binaries, Cygnus X-1 and GRS 1915+105. The latter two had no periodicities but did have data streams rich in variability at high temporal frequency. Study of these cases \cite{42} showed how the limitations come from both the intrinsic source variability at longer timescales (red noise), and from Poisson statistics, which give a white noise. The achievable accuracy was tens of microseconds. This might be used by itself or in combination with other forms of X-ray navigation. For example it could give a calibration to a system that was otherwise devoted primarily to pulsar-based navigation. 

\subsection{Lunar Occultations of the Crab Using RXTE (2011)}
\label{sec:4.4}

During the final months of \textit{RXTE} several attempts were made to schedule observations of lunar eclipses of the Crab Nebula viewed from orbit. These events recur at roughly 9 year intervals, or twice per cycle in the lunar precession period of $\sim$18.6 years. This was the fifth time since 1964 that the configuration had recurred. There had been rocket observations at the 1975 opportunity. The prior occasion in the life of \textit{RXTE} had gone unexploited but in 2011 the mission was extended a few months beyond its planned termination in part to enable the occultation observations. Proper planning required combining the relevant celestial mechanics with predictions of the \textit{RXTE} orbit, once again pointing to the intimacy between occultations and navigation. In general it was not known with confidence beforehand whether an occultation would or would not be captured but enough attempts were made that three occultation profiles with excellent statistics were obtained, each of them showing the abrupt drop at loss of the pulsar that had been sought in 1964 but which was then hidden by the noisy statistics of the small rocket payloads and the early detector designs. The successes were on two dates, 13 November 2011 and 11 Dec 2011. On the former date the yield consisted of one ingress event (in which count rate declines through the event as the Moon covers the Nebula) one egress (count rate rising) and in addition one emergence from occultation by the Earth atmosphere. The 11 December observation yielded only an ingress. These three should remain the best set of X-ray Lunar occultation profiles of any celestial source for years to come. A preliminary presentation \cite{43} of results below was made at a symposium honoring the end of the \textit{RXTE} mission. 

\begin{figure}
\includegraphics[width=4.5in]{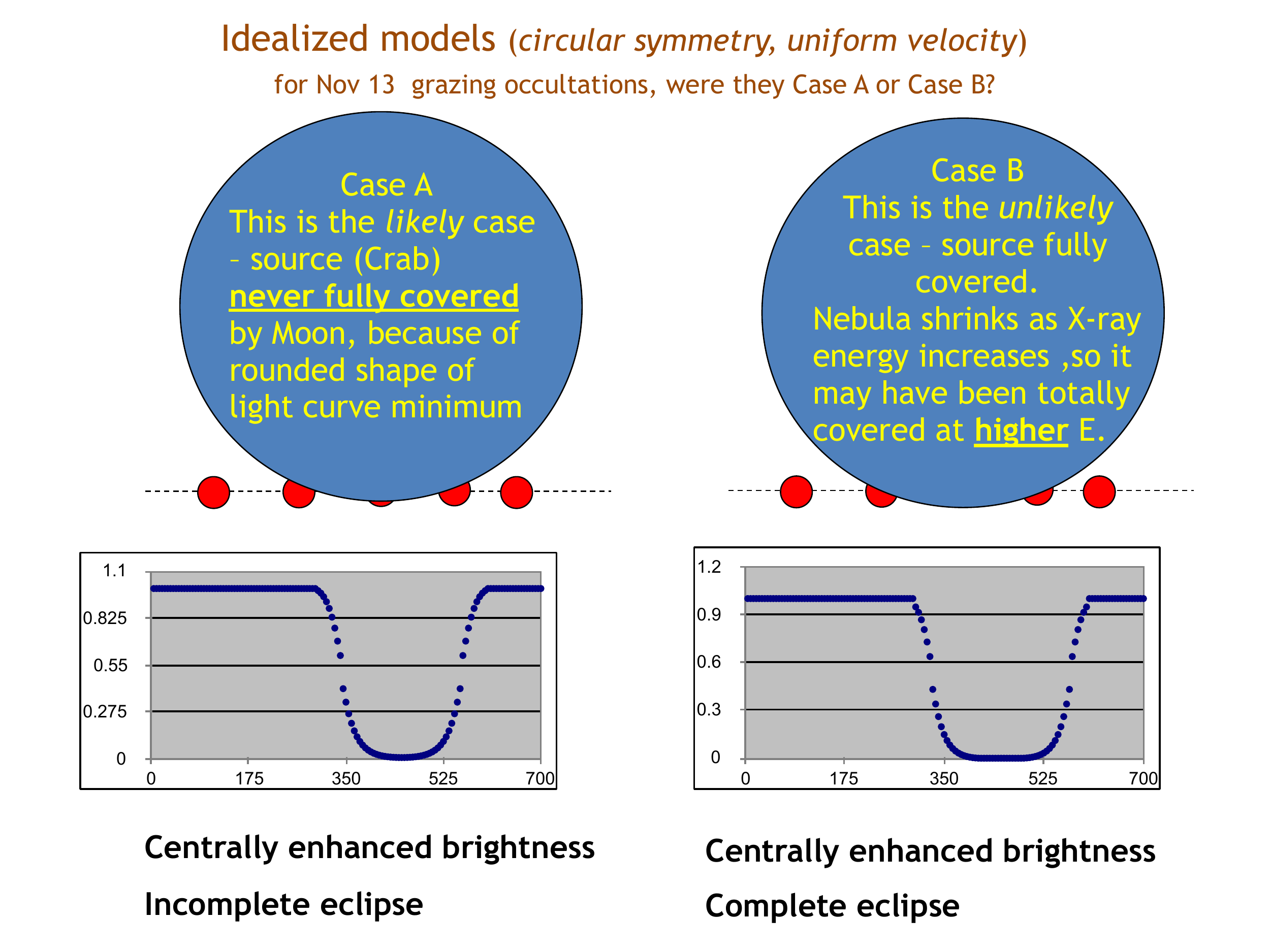}
\caption{Schematic contrasting idealized cases of near-limb lunar occultations of an extended source; blue disk represents Moon, red dot represents Crab Nebula.  Curves at bottom show relative coverage of Nebula by area fraction.}
\end{figure}

The time between ingress and egress is governed by whether the line joining ingress and egress points on the limb passes near the center of the lunar disk or comes closer to the poles. While the grazing case is less likely, that is what happened 13 November 2011. This occultation was so marginal that a small part of the Nebula may not have been covered, as is illustrated in Figure 10, which schematically compares the two possibilities in an idealized geometry where both Moon (blue) and Crab Nebula (red) are circles on the sky. The light curve for case B goes flat at the background rate, unlike case A.

\begin{figure}
\includegraphics[width=4.5in]{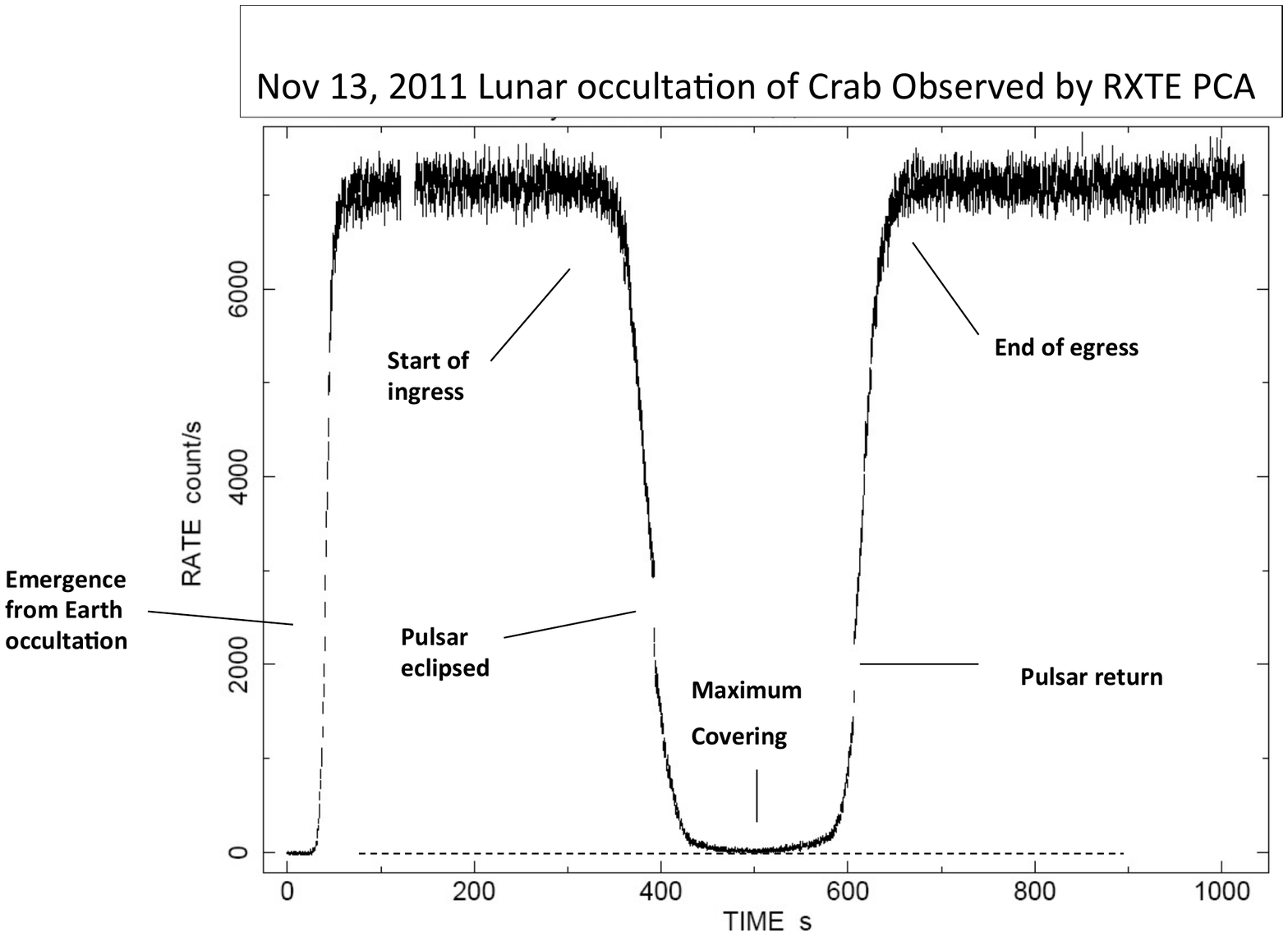}
\caption{Light curve of 13 Nov 2011 Occultation observed with \textit{RXTE} }
\end{figure}

\begin{figure}
\includegraphics[width=4.5in]{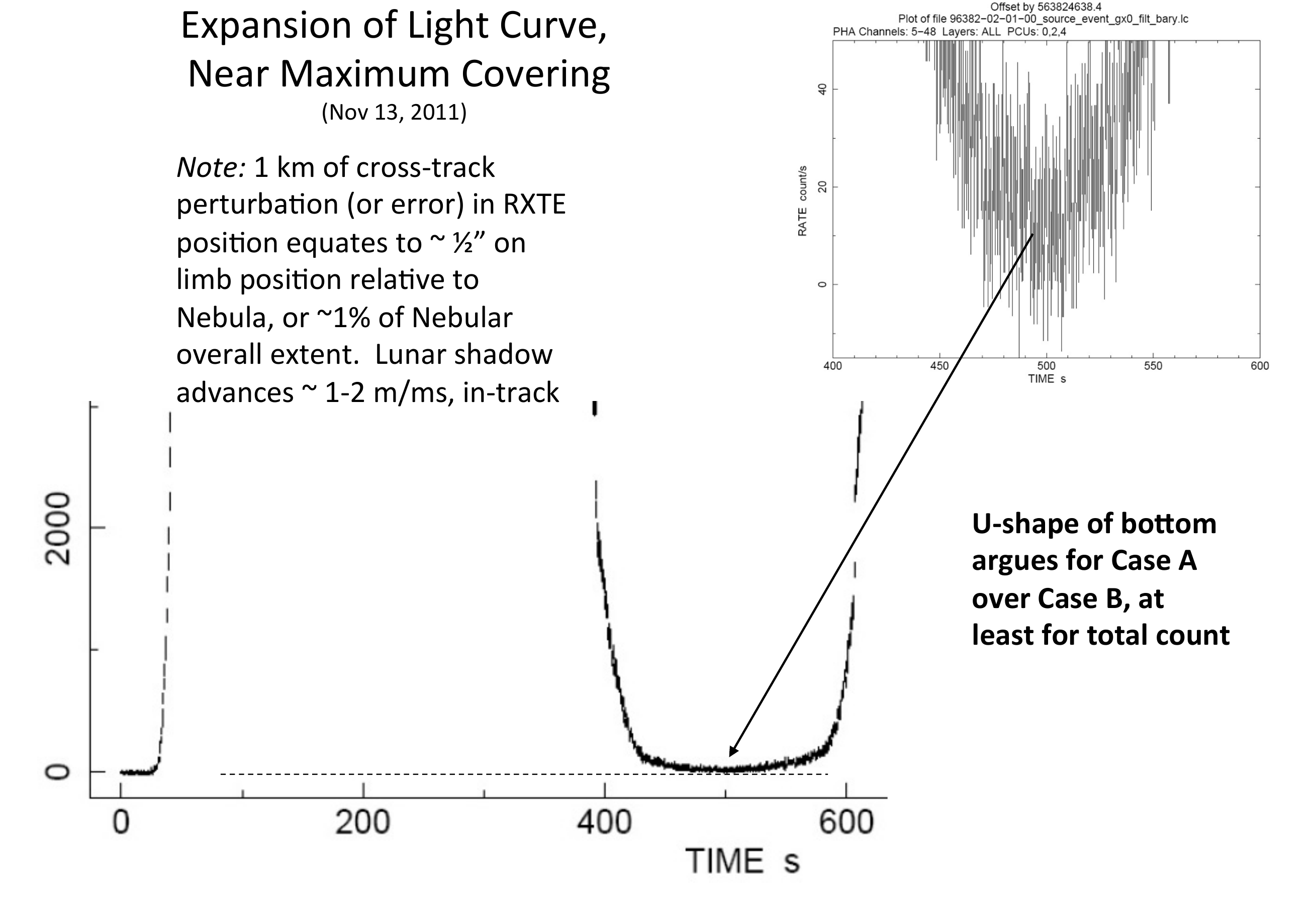}
\caption{Expansion of bottom portion of Figure 11 showing incomplete occultation. } 
\end{figure}

In case A the Crab is never fully covered because the chord (across the lunar disk joining points where the Nebula center crosses the lunar limb) comes so close to the edge that the maximum distance from the edge is less than the Nebular radius. In case B the Nebula is fully eclipsed. 

Whether the actual occultation is case A or B can be decided by close inspection of the occultation profile, as in Figure 11 with expansion to show detail in Figure 12. The rounded bottom, never going flat, is an indication of incomplete coverage. 

Consider the navigational ramifications: 1 km of cross-track perturbation (or error) in \textit{RXTE} position equates to $\sim$ 1/2 arcsec in the limb position relative to the Nebula, or ~1\% of the Nebular overall extent. The lunar shadow advances $\sim$ 1-2 m.a.s./ms, in-track The U-shape of bottom argues for Case A over Case B, at least for total count.

\begin{figure}
\includegraphics[width=4.5in]{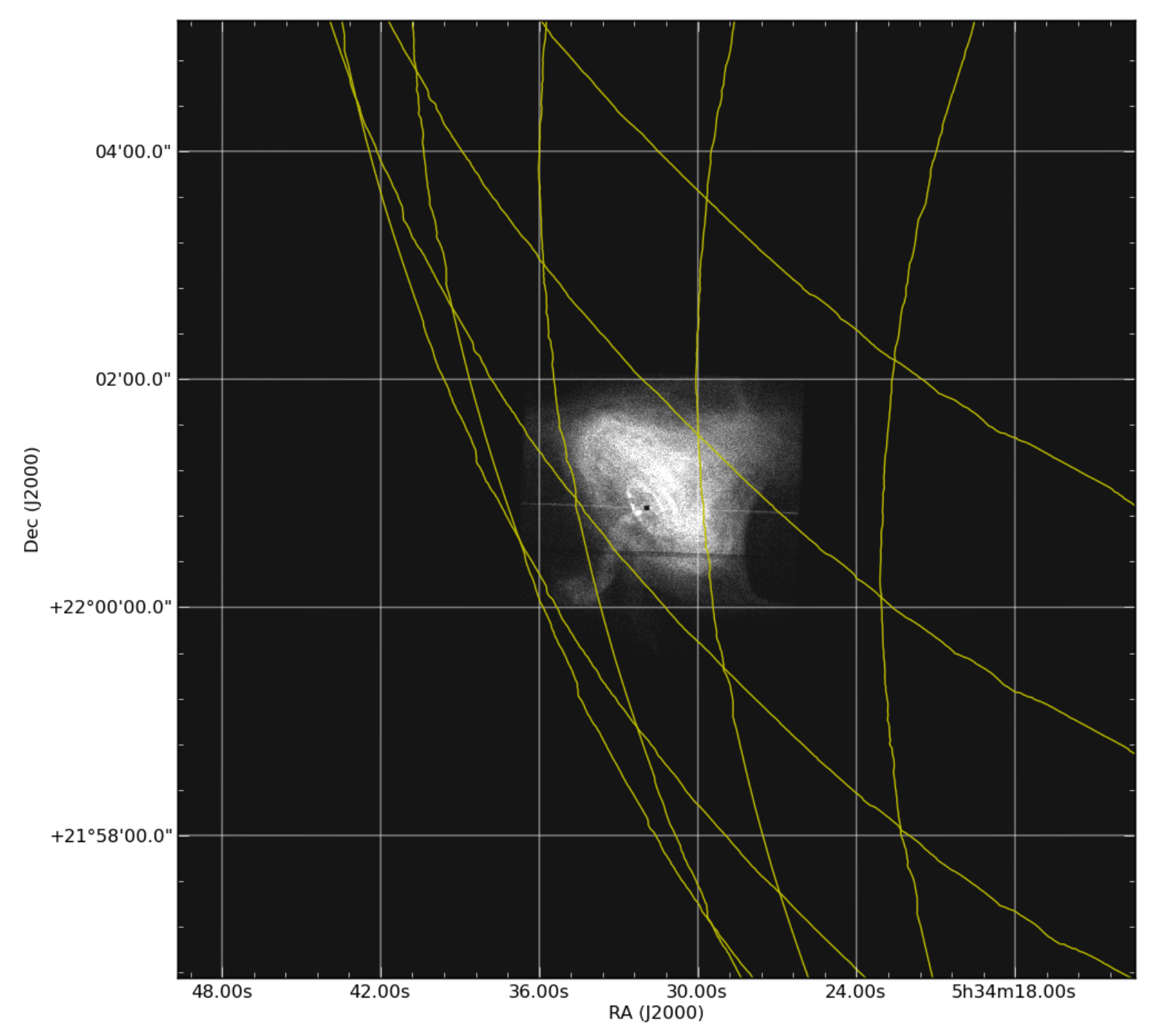}
\caption{A sequence of successive lunar limb profiles from 13 Nov 2011 occultation, including terrain, overlaid on Chandra X-ray image of the Crab.  Compare this figure with the right panel of Figure 1, which pertains to the 7 July 1964 sounding rocket flight.  }
\end{figure}

Figure 13 shows the proper way to do it, bringing in the exact lunar limb shape and its apparent motion relative to the Crab, by combining the motion of \textit{RXTE} and the Moon; overlaid on an X-ray image from Chandra. It shows that a small feature at the edge of the Nebula (sometimes called the Pipe) was apparently never covered. The X-ray image used is from a different X-ray energy band from that of the \textit{RXTE} response to the Nebula, although there is some overlap.

Detailed calculations using the lunar limb and nominal \textit{RXTE} trajectory may be used either for astrophysics or navigation, \textit{ i.e.} if the lunar limb and X-ray image are taken as truth then the eclipse profile provides a cross check on the \textit{RXTE} orbit, and could be used to constrain its error both in-track and cross-track.

\begin{figure}
\includegraphics[width=4.5in]{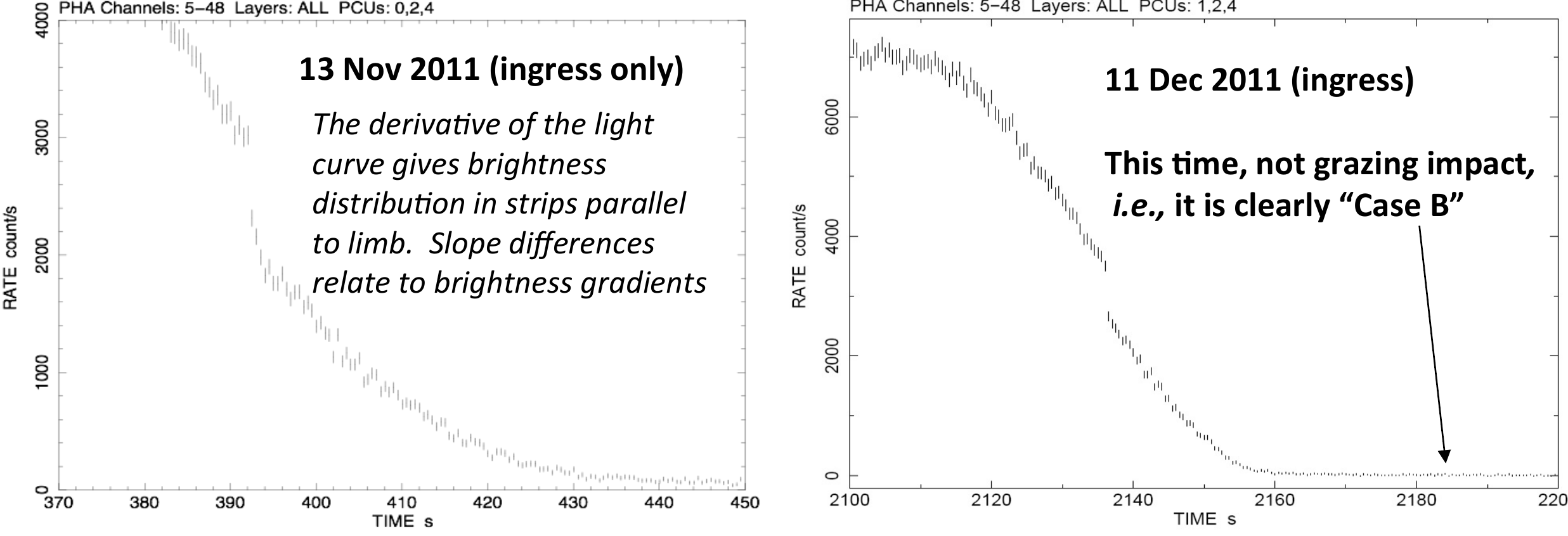}
\caption{Contrasting 13 Nov 2011 and 11 Dec 2011 occultations. }
\end{figure}

In contrast to 13 November 2011, the 11 December 2011 profile goes very flat at the bottom, confirming that in this instance the eclipse was complete. (Figure 14) The slopes of the three scan profiles (two from 13 November, one from 11 December) combined with the lunar and \textit{RXTE} orbits give information about the Nebular brightness distribution, measured in strips parallel to the advancing lunar limb in the three scans

\begin{figure}
\includegraphics[width=4.5in]{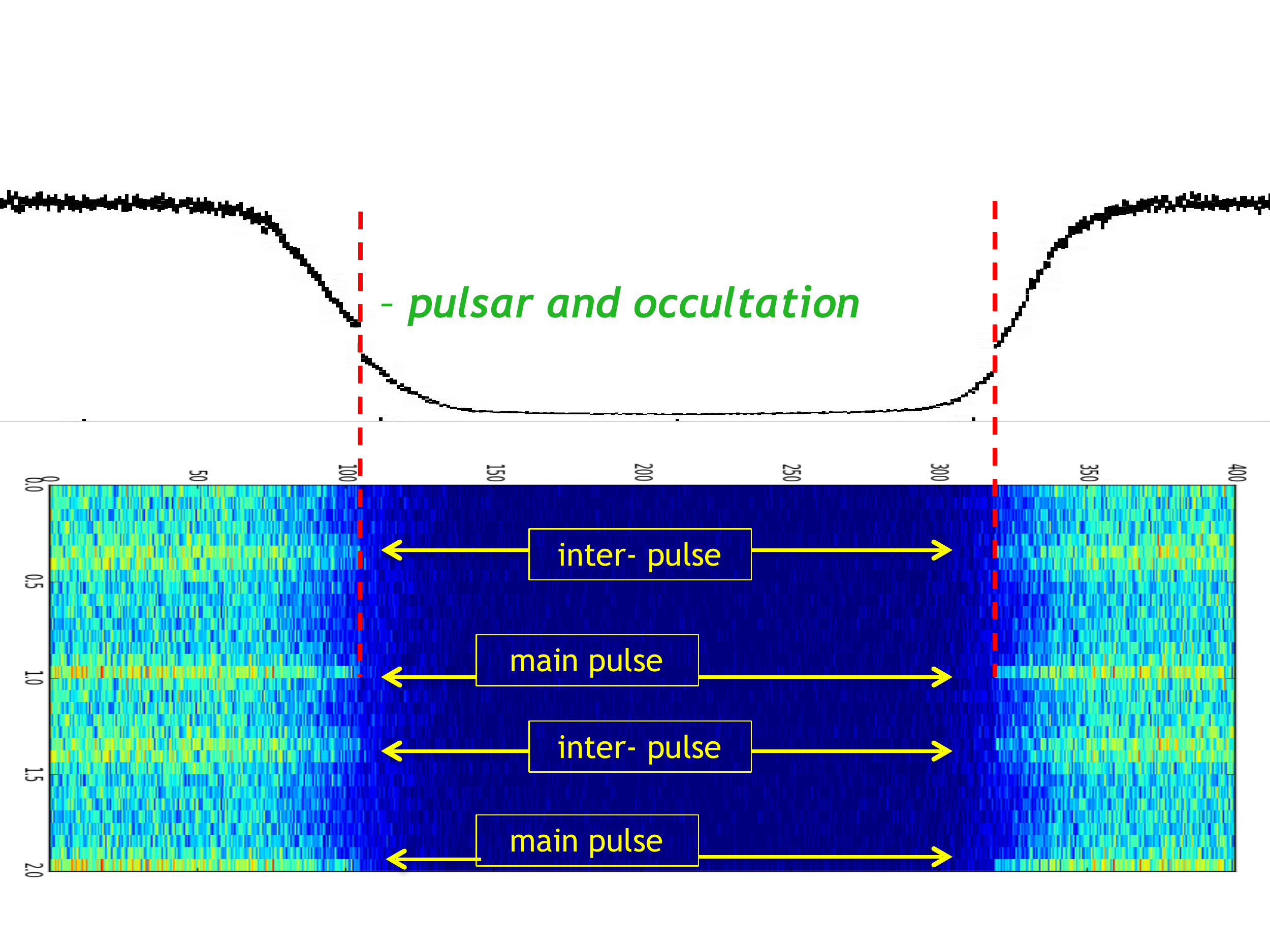}
\caption{Occultation of 13 November 2011 displayed to show pulsar phase.  Overall elapsed time increases to right.  In vertical direction one sees two pulsar cycles. }
\end{figure}

Finally, Figure 15 adds the additional information that is obtained by plotting the 13 November occultation and re-appearance count history as a function of pulsar phase. Each column of the figure is the count in a time bin equal to one cycle of the pulsar; the vertical direction gives the count in successive phase bins. One can see the pulsar blink out and return at the two times when the curve shows abrupt steps. This figure shows that on both sides the transition occurs during the part of the cycle between the end of the second pulse and return of the first. The pulsar information combined with an ephemeris is a cross check on the time as well as the position. This figure embodies the two themes that have run throughout the NRL program, pulsars and sharp transitions. 

\section{ Conclusions and Outlook}
\label{sec:5.0}

The NRL program in X-ray Navigation and allied topics has spanned five decades. It built upon early sounding rockets (1964), the lunar occultation of the Crab Nebula, the1969 discovery of the first X-ray pulsar), continued through the \textit{HEAO} A-1 survey, the USA program, and DARPA XNAV program, then went beyond those to exercises with \textit{RXTE} and participation in \textit{NICER/SEXTANT}. The program has explored timing approaches to navigation, using a variety of X-ray detector systems as X-ray technology advanced. The broad approach has explored attitude determination, position determination, timekeeping, and time transfer. 

X-ray navigation is now being assessed in space programs in a number of countries with a variety of approaches. Pulsar X-ray navigation (XNAV) is advancing towards a new demonstration with SEXTANT, using MSPs. Simulations and algorithm development are under way with NRL participation. NICER's science program will explore timing in general, with further progress expected navigational applications discussed here, including further study of transitions and relative navigation. If NICER lasts several years, it may observe Crab occultations by the Moon around 2020. 

X-ray astrophysics will evolve over coming decades in many ways, some of which will expand the relationship between navigation and angular information. An X-ray interferometer called MAXIM \cite{14} has been studied under NASA sponsorship. It will need very fine attitude knowledge/control. X-ray artificial occultations, aperture modulation or pulsar methods may find use. 

History has been a secondary aim of this chapter, the primary goal being to capture the XRNAV vision. The latter goal will now be brought to closure by looking at open questions that build on XRNAV accomplishments. It is plausible to imagine melding techniques studied into optimized X-ray navigation systems, either for flexibility in ambitious systems, or as niche solutions to particular problems. Candidate approaches must always be evaluated against non-X-ray solutions or with sensor fusion that includes X-rays.

\subsection{Occultations and Planetary Exploration}
\label{sec:5.1}

For long cruises between planets there is limited use for occultations. Rare, chance occultations might provide precise fiducials. Near planets, even chance occultations become common because the planet starts to subtend a large fraction of a steradian. This offers navigation referenced to the planet, especially accurate for bodies without atmospheres. It would be desirable to attempt this on a satellite in lunar orbit or perhaps on Mercury, for which a suitable mission will occur in the near future. Another chapter \cite{44} stresses the desirability of having a navigation sensor perform some mission function when it arrives at a destination, e.g., as optical sensors assist in surveys of encountered bodies. X-ray occultations could survey topography of airless planets or analyze density profiles of planets with atmospheres.

\subsection{Cold Start ---  the Minimum Requirement on Uplinked Information}
\label{sec:5.2}

Independence of a spacecraft of ground support relates to autonomy. Can an X-ray sensor system assign a spacecraft a position and time starting with some minimal complement of saved knowledge of source and ephemerides in memory?  Conjecturally, it might start by scanning to find Sco X-1 and the Crab and establishing orientation, then using the Crab Pulsar period to determine an approximate epoch. This initial position-plus-epoch estimate could be refined using eclipsing sources, then refined further using pulsars and their ephemerides until a limit is reached where further improvement depends upon receiving updates from Earth. For navigation in the outer Solar System and beyond, the Sun itself would become a coronal source whose brightness and position relative to other X-ray stars would carry navigational information. ``Cold start'' does not need to be implemented to have value as exploration of interdependencies of astronomical information and navigation. Characterizing the band over which binary eclipse centroid excursions extend is one scientific question in that would impact such a study.

\subsection{Relative Navigation or Time Transfer Using Optical and X-rays}
\label{sec:5.3}

The Crab Pulsar, as observed in the optical band, may be limited by timing noise but does have the advantage that it can be seen through the Earth atmosphere, with any time delay currently being miniscule. Thus a telescope on the ground can clock the main pulse of the Crab and provide a continuous timing update to a detector system on a satellite. There are possibilities that this same technique could be used for other sources. It is a variation on ``relative navigation'' but one where one of the satellites is the Earth itself. This is the case if data streams are cross correlated, which would not be exploiting the periodicity. One could also use pulsar methods, determining the TOA optically. 

\subsection{Cruise Navigation Using a Series of \textit{Post Facto} Determinations}
\label{sec:5.4}

The standard XNAV approach to position-determination on long (interplanetary) cruises is to extrapolate ephemerides for each pulsar into future time from the ephemeris epoch. It entails applying ephemerides outside fitted intervals, using them as truth until propagated errors grow too large. A different approach is to fit X-ray data to pulsar ephemerides only from intervals bracketed by observations Individual position-determinations are then running in arrears, somewhat behind in the actual satellite time (perhaps hours, perhaps a few days), but for their epochs they do not involve extrapolated ephemerides. This series of \textit{post facto} determinations can then itself be extrapolated, but now it is a different kind of extrapolation, namely dynamical extrapolation of a determined trajectory in space. Since use is limited to cruise phase, and since accelerations are smooth and limited in magnitude, such a propagated trajectory ought to give a respectable navigation solution. It is an open question whether this approach using a bright but noisy pulsars (the Crab and others) could compete with use of better clocks (MSPs) for concrete examples, always providing that use is limited to a cruise situation.

\begin{acknowledgement}
This work is supported by the Chief of Naval Research. \textit{NICER/SEXTANT} work at NRL is supported by NASA. 
%\fixcomment{Should I give NPR num?}
\end{acknowledgement}
\bibliographystyle{spphys}
\bibliography{Wood}
\end{document}